\documentclass[12pt,a4paper,final]{iopart}
\usepackage{amsfonts}
\usepackage{eurosym}
\usepackage{iopams}
\usepackage{graphicx}
\usepackage[breaklinks=true,colorlinks=true,linkcolor=blue,urlcolor=blue,citecolor=blue]{hyperref}
\usepackage{hyperref}
\usepackage{subfigure}
\usepackage{caption}
\usepackage{bm}
\usepackage{verbatim}
\usepackage{epstopdf}
\usepackage{float}
\usepackage{tikz}

\providecommand{\U}[1]{\protect\rule{.1in}{.1in}}

\begin{document}

\title[Sudden Decoherence Transitions for Quantum Discord]{Sudden
Decoherence Transitions for Quantum Discord}
\author{Hyungjun Lim, Robert Joynt}

\begin{abstract}
We investigate the disappearance of discord in 2- and multi-qubit systems
subject to decohering influences. We formulate the computation of quantum
discord and quantum geometric discord in terms of the generalized Bloch
vector, which gives useful insights on the time evolution of quantum
coherence for the open system, particularly the comparison of entanglement
and discord. We show that the analytical calculation of the global geometric
discord is NP-hard in the number of qubits, but a similar statement for
global entropic discord is more difficult to prove. We present an efficient
numerical method to calculating the quantum discord for a certain important
class of multipartite states. In agreement with previous work for 2-qubit
cases, (Mazzola \textit{et al.}, Phys. Rev. Lett. 104, 200401 (2010)), we
find situations in which there is a sudden transition from classical to
quantum decoherence characterized by the discord remaining relatively robust
(classical decoherence) until a certain point from where it begins to decay
quickly whereas the classical correlation decays more slowly (quantum
decoherence). However, we find that as the number of qubits increases, the
chance of this kind of transition occurring becomes small.
\end{abstract}


\address{Department of Physics, University of Wisconsin-Madison, Madison, WI
53706, US} 
\ead{\mailto{\color{blue}hlim29@wisc.edu},
\mailto{\color{blue}rjjoynt@wisc.edu}}


\vspace{2.0075pc} \noindent \textit{Keywords}: quantum discord, quantum
entanglement, quantum decoherence, sudden transition, injective tensor norm,
NP-hard problem

\submitto{\JPA}

%\maketitle

\section{Introduction}

Entanglement is a unique property of composite quantum systems. It is known
to be essential for certain quantum communication protocols \cite%
{cleve_substituting_1997, prevedel_entanglement-enhanced_2011}, and it is
generally felt to be an essential resource of the exponential speedup of
quantum algorithms \cite{jozsa_role_2003, vidal_efficient_2003}. For
instance, the Deutsch-Josza algorithm (DJA) (the simplest of all quantum
algorithms) \cite{deutsch_rapid_1992} is designed to figure out if a given
function is balanced or not. \ Entanglement is unavoidable in the DJA as the
number $n$ of qubit increases \cite{bru_multipartite_2011}, since the total
number of balanced functions scales doubly exponentially: $%
B(2^{n},2^{n-1})\sim2^{2^{n}}$, while the number of separable states scales
as $2^{n}$. Thus, in order to represent all the balanced function with $n$
qubits, we inevitably introduce entanglement.

On the other hand, there have been questions concerning whether entanglement
is the only resource of the power of quantum algorithms \cite%
{lloyd_quantum_1999, meyer_sophisticated_2000}. \ The DJA for the 2- and 3-
qubit cases have an advantage over classical algorithm but they do not
involve much more than simple interference, not usually thought of as
quantum correlation. \ Quantum discord has been introduced as another type
of quantum correlation \cite{ollivier_quantum_2001} that can be present even
in separable states. \ It might be an additional resource of quantum
algorithms giving computational advantages over classical calculation. \
Evidence for this point of view comes from the existence of an algorithm
(DQC1) that computes the trace of a unitary matrix. \ It uses only one pure
qubit and exhibits discord but little entanglement \cite%
{datta_entanglement_2005, datta_quantum_2008, passante_measuring_2012}, and
still appears to be more powerful than classical algorithms for the same
problem. Since quantum discord is more robust against decoherence than
entanglement there is the hope that quantum algorithms dependent only on
quantum discord might be more feasible to implement physically than those
dependent on the more fragile entanglement. \ The true physical resource of
quantum computation still has many open issues.

Quantum discord was originally suggested as a quantum measure of
correlations between two systems that is analogous to the classical notion
of mutual information between two probability distributions in classical
information theory. \ It can be given an operational meaning as well \cite%
{cavalcanti_operational_2011, gu_observing_2012}. \ The original quantum
discord was defined only for bipartite states. \ For 2-qubit systems, it has
been shown that quantum discord can undergo a transition between quantum and
classical decoherence as a function of time on certain physically plausible
dynamical trajectories\cite{maziero_classical_2009, mazzola_sudden_2010,
bellomo_compagno_franco_ridolfo_savasta_2011, aaronson_franco_adesso_2013}.

Our chief focus in this paper is on multipartite systems. \ For this
purpose, the global quantum discord \cite{rulli_global_2011} is defined
using distance measures to the nearest classical state. The distance can be
defined either by the relative entropy measure \cite{modi_unified_2010}, or
by means of the metric derived from the Hilbert-Schimdt inner product \cite%
{luo_geometric_2010, xu_geometric_2012-1, xu_geometric_2012}. We will use
the latter form in this work, and it is called geometric global quantum
discord (GGQD). It facilitates the calculations by removing the
nonpolynomial log function from the definition, but at the cost of muddying
the information-theoretic interpretation.

The structure of the paper is as follows. The next section briefly
introduces the various measures of discord and elucidates their physical
meaning. The third section gives an interpretation of quantum discord in
terms of Bloch vector geometry, a proof that computing GGCD is NP-hard in
the number of qubits, and also treats the problem of 2,3, and N qubits from
an algebraic point of view. The fourth section describes a relatively simple
heuristic method to calculate GGQD for certain highly symmetric but
important classes of multiqubit states, which in turn suggests a condition
to obsesrve sudden transition between classical and quantum decoherence \cite%
{maziero_classical_2009, mazzola_sudden_2010} in the multiqubit case. This
condition appears to be very stringent, meaning that such transitions are
not likely to be observed on physical trajectories.

\section{Measures of quantum correlation}

A good measure of the classical correlation between two probability
distributions $\{p_{1,i}\},\{p_{2,j}\}$ with $\{p_{12,ij}\}$ as their joint
distribution is given by the mutual information $I_{classical}=H(\{p_{1,i}%
\})+H(\{p_{2,j}\})-H(\{p_{12,ij}\})$ where $H(\{p_{k}\})\equiv-\sum_{k}p_{k}%
\log_{2}(p_{k})$ is Shannon entropy. Note that $H(\{p_{12,ij}\})=H(\{p_{1,i}%
\})+H(\{p_{2,j}\})$ when two distributions are independent so that $%
I_{classical}=0$. $I_{classical}$ can also be written as 
\begin{eqnarray}
& I_{classical}=H(\{p_{1,i}\})+H(\{p_{2,j}\})-H(\{p_{12,ij}\})  \nonumber \\
& =H(\{p_{1,i}\})-(H(\{p_{12,ij}\})-H(\{p_{2,j}\})) \\
& =H(\{p_{1,i}\}) -\left(-\sum_{ij}p_{12,ij}\log_{2}\left( p_{12,ij}\right)
+\sum_{ij}p_{12,ij}\log_{2}(p_{2,i})\right)  \nonumber \\
& =H(\{p_{1,i}\})-\left( - \sum_{ij}p_{2,j}\left( \frac{p_{12,ij}}{p_{2,j}}%
\right) \log_{2}(\frac{p_{12,ij}}{p_{2,j}})\right)  \nonumber \\
& =H\left( \left\{ p_{1,i}\right\} \right) -\sum_{j}p_{2,j}H\left( \left\{
p_{1|2=j,i}\right\} \right)
\end{eqnarray}
Here the conditional probability $p_{1|2=j,i}$ is the probability of event $%
i $ in system 1, given that event $j$ has been measured in system 2. \ 

This process can be viewed as the partial elimination of uncertainty of
system 1 from knowledge of system 2. The classical correlation between two
quantum systems can be defined in the light of these equations. In the
quantum case, the probability distribution is replaced by the density
operator $\rho$, and the Shannon entropy $H$ is replaced by the von Neumann
entropy $S(\rho)\equiv H(\{\lambda_{\rho,k}\})=-\sum_{k}\lambda_{\rho,k}$log$%
_{2}(\lambda_{\rho,k})$ where ${\lambda_{\rho,k}}$ are singular values of $%
\rho.$ \ We see that the classical conditional probability distribution $%
\{p_{1|2=j,i}\}$ is analogous to $\rho_{1|j}$, the state of system 1 after a
projective measurement $\Pi _{j}^{(2)}$ is applied on the second system,
thus a reasonable definition of the classical correlation is then given by 
\[
J_{\{\Pi_{j}^{(2)}\}}=S(\rho_{1})-\left( \sum_{j}p_{j}S(\rho_{1|j}\otimes|j%
\big>\big<j|)\right) . 
\]

But this quantity still depends on the measurement choice (the big
difference between quantum and classical mechanics). \ \ If we make the best
choice, then we finally find the \textit{optimized} classical correlation $%
J_{1|2}$ between two quantum systems: 
\begin{equation}
J_{1|2}=\max_{\{\Pi_{j}^{(2)}\}}J_{\{\Pi_{j}^{(2)}\}}.
\end{equation}
Quantum discord is defined as the difference of the total correlation $%
I(\rho)=S(\rho_{1})+S(\rho_{2})-S(\rho)$, and the optimized classical
correlation \cite{ollivier_quantum_2001}, 
\[
D_{1|2}(\rho)=I(\rho)-J_{1|2}(\rho) 
\]

This original definition of quantum discord has certain drawbacks. \ It is
asymmetric between systems 1 and 2 so there is ambiguity of which to choose.
It is defined only for bipartite systems so an extension of the definition
to multipartite systems is needed. \ It involves an optimization which can
be challenging. \ \ 

These problems can be alleviated somewhat by noting that $D_{1|2}(\rho )$
vanishes in quantum-classical states, defined as all states of the form 
\[
\{\sum_{i}p_{i}\rho _{i}\otimes |i\big>\big<i|;p_{i}\leq 0,\big<i|j\big>%
=\delta _{i,j}\}. 
\]%
We can then define the geometric discord of a given state as the distance
from that state to the nearest quantum-classical state. \ D. Girolami 
\textit{et al.}\ obtained simplified analytic conditions for the optimal
value for the 2-qubit case \cite{girolami_quantum_2011}. \ This definition
is still asymmetrical and requires numerical parameter scans. \ 

Rulli \textit{et al.} suggested that one can make quantum discord into a
symmetrical, multipartite by defining the global quantum discord\cite%
{rulli_global_2011}. The original quantum discord can also be written as 
\[
D_{1|2}=\min_{\{\Pi^{(2)}_{j}\}}\big[I(\rho)-S(\Pi^{(2)}(\rho)\parallel%
\rho_{1}\otimes tr_{1}(\Pi^{(2)}(\rho)))\big]
\]
where 
\[
\Pi^{(2)}(\rho)\equiv\sum_{j}(I\otimes\Pi_{j}^{(2)})\rho(I\otimes%
\Pi_{j}^{(2)})=\sum_{j}p_{j}\rho_{1|j}\otimes|j\big>\big<j| 
\]
and 
\[
S(\hat{x}\parallel\hat{y})=-tr(\hat{x}\log\hat{y})-S(\hat{x}) 
\]
using $S(\sum_{j}p_{j}\rho_{1|j}\otimes|j\big>\big<j|)=S(\sum_{j}p_{j}|j\big>%
\big<j|)+\sum_{j}p_{j}S(\rho_{1|j}\otimes|j\big>\big<j|)$\cite%
{modi_unified_2010, xu_geometric_2012}. Motivated by this observation, the
global quantum discord is defined by expanding the measurement $\Pi^{(2)}$
on the second part, to the measurement $\Pi^{(1,2)}$ on both parts.

\begin{eqnarray}
&&D_{G}=\min_{\{\Pi _{i_{1}}^{(1)}\},\{\Pi _{i_{2}}^{(2)}\}}\bigg(I(\rho )-S%
\big(\Pi ^{(1,2)}(\rho )\parallel tr_{2}(\Pi ^{(1,2)}(\rho ))\otimes
tr_{1}(\Pi ^{(1,2)}(\rho ))\big)\bigg)  \nonumber
\label{eq:global quantum discord} \\
&=&\min_{\{\Pi _{i_{1}}^{(1)}\},\{\Pi _{i_{2}}^{(2)}\}}\big(I(\rho )-I(\Pi
^{(1,2)}(\rho ))\big)
\end{eqnarray}%
where 
\[
\Pi ^{(1,2)}(\rho )\equiv \sum_{i_{1},i_{2}}(\Pi _{i_{1}}^{(1)}\otimes \Pi
_{i_{2}}^{(2)})\rho (\Pi _{i_{1}}^{(1)}\otimes \Pi
_{i_{2}}^{(2)})=\sum_{i_{1},i_{2}}p_{i_{1},i_{2}}|i_{1}\big>\big<%
i_{1}|\otimes |i_{2}\big>\big<i_{2}| 
\]%
and $I(\rho )=S(\rho \parallel tr_{2}(\rho )\otimes tr_{1}(\rho ))$ was used 
\cite{modi_unified_2010}. Rulli \textit{et al.}\cite{rulli_global_2011}
performed analytical calculations for the 2-qubit Werner-GHZ state and X.
Jianwei applied it to a multiqubit Werner-GHZ state \cite%
{xu_geometric_2012-1}. Furthermore, we can make an observation that if the
state is in the set $Cl$ of classical states defined by 
\[
\{\sum_{i_{1},i_{2}}p_{i_{1},i_{2}}|i_{1}\big>\big<i_{1}|\otimes |i_{2}\big>%
\big<i_{2}|;p_{i_{1},i_{2}}\leq 0,\big<i_{1}|j_{1}\big>=\delta
_{i_{1},j_{1}},\big<i_{2}|j_{2}\big>=\delta _{i_{2},j_{2}}\} 
\]%
then $D_{G}=0$. \ In this spirit, geometric global quantum discord (GGQD) is
defined as the distance to the nearest classical state\cite%
{xu_geometric_2012-1}. 
\[
D_{GG}(\rho )\min_{\chi \in Cl}~\left\vert \rho -\chi \right\vert ^{2}, 
\]%
where we use the Hilbert-Schmidt metric: 
\[
\left\vert \rho -\chi \right\vert ^{2}=\Tr\left[ \left( \rho -\chi \right)
^{\dagger }\left( \rho -\chi \right) \right] 
\]%
For two qubit states, the above definition was scaled by the factor of $%
\frac{1}{2}.$ \ With this, then $1\geq 2D_{GG}\geq D_{G}$ \cite%
{nguyen_topology_2013}. The Hilbert-Schmidt metric has many advantages: it
is relatively easy to calculate, it can be interpreted in terms of
measurements of Pauli operators, and since it is related to the Euclidean
metric, it has a simple geometrical interpretation, which allows one to use
more powerful optimization methods . \ By contrast, the definition of GQD
involves logarithms, which complicates the optimization. As GGQD removes the
logarithm, we will instead use it, even though its interpretation as a
resource for quantum information processing is less clear.

It was shown \cite{xu_geometric_2012} that GGQD for n qubits state $\rho $
can be evaluated as follows: 
\begin{equation}
D_{GG}(\rho )=\frac{1}{2^{n}}(\sum_{a}n_{a}^{2}-\max_{\{\vec{\Theta _{i}}%
\}}(\sum_{a}n_{a}\Pi _{i=1}^{n}\Theta _{i,a_{i}})^{2})
\label{eq:geometric global quantum discord}
\end{equation}%
where measurement on each qubit is along the qubit direction $\vec{\Theta}%
_{i}$ so that $\Pi _{m=1}^{(i)}=|\vec{\Theta}_{i}\big>\big<\vec{\Theta}_{i}|$%
, with $|\vec{\Theta}_{i}\big>=\cos (\frac{\theta (\vec{\Theta}_{i})}{2})|0%
\big>+e^{\imath \phi (\vec{\Theta}_{i})}\sin (\frac{\theta (\vec{\Theta}_{i})%
}{2})|1\big>$ where $\theta (\vec{\Theta}_{i})$ and $\phi (\vec{\Theta}_{i})$
are the polar and azimuthal angles of the unit vector $\vec{\Theta}_{i}$,
and $\Pi _{m=2}^{(i)}=|\vec{\Theta}_{i}^{\prime }\big>\big<\vec{\Theta}%
_{i}^{\prime }|$ with $|\vec{\Theta}_{i}^{\prime }\big>=\sin (\frac{\theta (%
\vec{\Theta}_{i})}{2})|0\big>-e^{\imath \phi (\vec{\Theta}_{i})}\sin (\frac{%
\theta (\vec{\Theta}_{i})}{2})|1\big>$.

\section{Calculation Of Quantum Discord}

\subsection{Bloch vector for generalized Bell States}

We represent the density matrix $\rho$ for an $N$-qubit system using the
generalized Bloch vector $n_{a}:$%
\begin{equation}
n_{a}=Tr~(\rho O_{a})  \label{eq:comp}
\end{equation}
where the subscript $a$ labels the generators $O_{a}$ of SU$\left(
2^{N}\right) ,$ taken as tensor products of the Pauli matrices $\sigma
_{0}=I,\sigma_{1,2,3}=\sigma_{x,y,z}$. \ Thus $a\in\{0,3\}^{N},$ \textit{i.e.%
}, $a$ is an $N$-digit base-4 number. \ The trace condition on $\rho$ gives $%
n_{\vec{0}}=1$ always, so we omit the index $a=\vec{0}$ in what follows. \
Inverting Eq.(\ref{eq:comp}) gives 
\[
\rho=\frac{1}{2^{n}}(I_{n}+\sum_{a=00..01}^{4^{N}-1}n_{a}O_{a}) 
\]
The $n_{a}$ are the real components of a $(4^{N}-1)$-dimensional vector,
which can be thought of as a generalization of the Bloch vector. \
Positivity requirements on $\rho$ lead to a state space $\mathcal{M}$ that
is a subset of $\mathbb{R}^{4^{N}-1}.$ For given $N,\mathcal{M}$ is compact
and convex, but its surface has a complicated shape \cite%
{byrd_characterization_2003, zhou_disappearance_2011}.

As mentioned in the previous section, GQD introduces a measurement on the
all parts of the system and in order to calculate $I(\Pi ^{(1,2)}(\rho ))$
in Eq.(\ref{eq:global quantum discord}), we need the eigenvalue spectrum of
the state $\Pi ^{(1,2)}(\rho )$ after the measurement is applied. In this
paper, we focus on a linear subspace of $\mathcal{M}$ (more precisely, the
intersection of a $3^{N}$-dimensional linear subspace of $\mathbb{R}%
^{4^{N}-1}$ with $\mathcal{M}$.) \ This subspace is defined by the condition
that $n_{a}=0$ if any digit of $a$ is zero. \ Then the marginal state of any
subsystem is maximally mixed: if we take the partial trace over any subset
of the qubits, the remaining marginal state is a state whose density matrix
is proportional to identity matrix. \ From that point of view, these $N$%
-qubit states can be thought of as generalizations of the 2-qubit Bell
states. \ Unlike the 2-qubit Bell states, these states are not known to be
maximally entangled in any sense, but they are good candidates for highly
entangled states with relatively few parameters.

We now perform measurements along the angles given by $\{\vec{\Theta}_{i}\}$%
, a set of unit vectors, given $\rho $ in this described subspace. The
eigenvalues of $\rho $ after the measurements are what we need for the
calculation of the discord. There are only two distinct eigenvalues given by 
$2^{-N}[1\pm C(\{\vec{\Theta}_{i}\})]$, and each is $2^{N-1}-$fold
degenerate. Here $C(\{\vec{\Theta}{_{i}}\})\equiv \sum_{\vec{a}%
}n_{a_{1},\cdots ,a_{N}}\Theta _{1,a_{1}}\cdots \Theta _{N,a_{N}}$. \ The
global quantum discord and geometric global quantum discord are, from Eq.(%
\ref{eq:global quantum discord}) and Eq.(\ref{eq:geometric global quantum
discord}), 
\begin{eqnarray}
D_{G}(\rho ) &=&I(\rho )+\min_{\{\vec{\Theta}_{i}\}}H(\{\frac{1}{2}(1+C),%
\frac{1}{2}(1-C)\})-(N-1)  \nonumber
\label{eq:global quantum discord for special case} \\
&=&I(\rho )+H(\{\frac{1}{2}(1+\max_{\{\vec{\Theta}_{i}\}}C),\frac{1}{2}%
(1-\max_{\{\vec{\Theta}_{i}\}}C)\})-(N-1)
\end{eqnarray}%
and 
\begin{eqnarray}
D_{GG}(\rho ) &=&\frac{1}{2^{n}}(\sum_{a}n_{a}^{2}-\max_{\left\{ \vec{\Theta}%
_{i}\right\} }(\sum_{a}n_{a}\Pi _{i=1}^{n}\Theta _{i,a_{i}})^{2})  \nonumber
\label{eq:geometric global quantum discord for special case} \\
&=&\frac{1}{2^{n}}(\sum_{a}n_{a}^{2}-\max_{\{\vec{\Theta}_{i}\}}C^{2})
\end{eqnarray}%
The Shannon entropy function $H$, the measure of randomness, is larger when
the probability is more evenly distributed, so the optimization problem for
both GQD and GGQD becomes the task of finding maximum $C$. \ $C$ is the
contraction of the tensors $n$ and the product of the 1-tensors $\vec{\Theta}%
_{i}$ \ For two qubits, $\max_{\left\{ \vec{\Theta}_{i}\right\} }C$ is the
operator norm of the matrix $n$ whose calculation is essentially an
eigenvalue problem, but even for three qubits, an analytic expression for $%
\max_{\{\vec{\Theta}_{i}\}}\sum_{ijk}n_{ijk}\Theta _{1,i}~\Theta
_{2,j}~\Theta _{3,k}$ is not available in general. \ 

\subsection{NP-hardness of quantum discord calculation}

\begin{eqnarray}  \label{eq:injective tensor norm}
\max_{\{\vec{\Theta}_{i}\}}C(\{\vec{\Theta}_{i}\})\equiv\max_{\{\vec{\Theta}%
_{i}\}}\sum_{a_{1}\cdots a_{N} \in \{1,2,3\}^{\otimes N}}n_{a_{1}\cdots
a_{N}} \Pi_{i=1}^{N}\Theta_{i,a_{i}}
\end{eqnarray}
is called the injective tensor norm of the tensor $n_{a}$ with $N$ indices\cite{harrow_montanaro_2012}.
There are various ways to define a norm on tensors, the injective tensor
norm is the most straightforward generalization of the operator norm, to
which it reduces when $N=2$. Physically one may think of it as the ground
state energy of a classical spin glass with unit vector spins, with
arbitrary $N$-spin interactions and a Hamiltonian $H=-\sum_{a}n_{a}~%
\Theta_{1,a_{1}}\cdots ~\Theta_{N,a_{N}}$.

For the general state, the geometric global quantum discord (GGQD, Eq.(\ref%
{eq:geometric global quantum discord})) involves the calculation of $\max_{\{%
\vec{\Theta_{i}}\}}(\sum_{a\in\{0,\cdots,3\}^{\otimes
N}}n_{a}\Pi_{i=1}^{N}\Theta_{i,a_{i}})$. It is similar to the injective
tensor norm Eq.(\ref{eq:injective tensor norm}), but now $n_a$ is not
necessarily zero when some digit of $a$ is zero. It can be shown that GGQD
calculation is NP-hard by converting an NP-hard MAX-k-SAT problem to GGQD
calculation.

In a MAX-k-SAT problem, we have M clauses, $\{C_{i}:i=1,\cdots M\}$ where
each clause $C_{i}$ involves k boolean variables, $V_{i}\equiv%
\{v_{c(i,j)}:j=1,\cdots k,c(i,j)\in\mathbb{N}\}$ and the whole problem
involves N variables, $V\equiv\cup_{i}V_{i}=\{v_{i}:i=1,\cdots,N\}$. For
each clause $C_{i}$, there is a set of assignments for variables in $V_{i}$
that satisfy $C_{i}$, $A_{i}=\{A_{i,l}:A_{i,l}=\{v_{c(i,j)}=a_{c(i,j),l}%
\},a_{c(i,j),l}\in\{T,F\},l=1,\cdots,|A_{i}|\}$. The problem is to find an
assignment $A^{\ast}=\{v_{i}=a_{i}^{\ast}\}$ of the all variables that
maximizes the number of the satisfied clauses: $A^{\ast}=\arg\max_{A}$k-SAT$%
(A)$ where k-SAT$(A)$ is the number of clauses.

To make the connection between the GGQD calculation and MAX-k-SAT problem,
define a measurement angle $\vec{\Theta}_{i}$ for each variable $v_{i}$ in
V. For an assignment of a variable, we define the corresponding measurement
angle such that $\vec{\Theta}(T)=(0,0,1)^{T}$ and $\vec{\Theta}%
(F)=(0,0,-1)^{T}$. For each clause $C_{i}$, define the corresponding
Hamiltonian $H_{i}$ such that $H_{i}$ has the value of $-1$ for the
assignments that satisfy the clause and $0$ otherwise: $H_{i}=-\sum_{l}%
\Pi_{j=1}^{k}\{\frac{1}{2}(1+\Theta(a_{c(i,j),l})_{z}\Theta_{j,z})\}$. The
total Hamiltonian is the sum of all $H_{i}$ and $n_a$ is defined
accordingly: $H_{tot}=\sum_{i=1}^{N}H_{i}=-\sum_{a\in\{0,\cdots,3\}^{\otimes
N}}n_{a}\Pi_{i=1}^{N}\Theta_{i,a_{i}}$. The measurement angles $\{\vec{\Theta%
}_{i}^{\ast}\equiv\vec{\Theta}(a_{i}^{\ast}):i=1,\cdots,N\}$ corresponding
to $A^{\ast}$ also make $H_{tot}$ to have the lowest value.

As a simple example, for $C_{1}=v_{1}\vee v_{2}$, $C_{2}=v_{1}\vee\lnot
v_{2} $, we have 
\begin{eqnarray*}
H_{1}&=-\frac{(1+\Theta_{1,z})}{2}\frac{(1+\Theta_{2,z})}{2} \\
&-\frac{(1+\Theta_{1,z})}{2}\frac{(1-\Theta_{2,z})}{2}-\frac{(1-\Theta_{1,z})%
}{2}\frac{(1+\Theta_{2,z})}{2}
\end{eqnarray*}
and 
\begin{eqnarray*}
H_{2}&=-\frac{(1+\Theta_{1,z})}{2}\frac{(1+\Theta_{2,z})}{2} \\
&-\frac{(1+\Theta_{1,z})}{2}\frac{(1-\Theta_{2,z})}{2}-\frac{(1-\Theta_{1,z})%
}{2}\frac{(1-\Theta_{2,z})}{2}
\end{eqnarray*}
so that 
\begin{eqnarray*}
H_{tot}&=-2\frac{(1+\Theta_{1,z})}{2}\frac{(1+\Theta_{2,z})}{2}-2\frac{%
(1+\Theta_{1,z})}{2}\frac{(1-\Theta_{2,z})}{2} \\
&-\frac{(1-\Theta_{1,z})}{2}\frac{(1+\Theta_{2,z})}{2}-\frac{(1-\Theta_{1,z})%
}{2}\frac{(1-\Theta_{2,z})}{2} \\
&=-\frac{3}{2} - \frac{1}{2}\Theta_{1,z}
\end{eqnarray*}
and the ground state is obtained with $\vec{\Theta}_{1}=\vec{\Theta}(T)$
which means $\Theta_{2,z}$ can either be $+1$ or $-1$. It corresponds to the
assignments of $\{v_{1}=T,v_{2}=T\}$ or $\{v_{1}=T,v_{2}=F\}$.

Therefore, by being able to calculate $\max_{\{\vec{\Theta_{i}}%
\}}(\sum_{a\in\{0,\cdots,3\}^{\otimes N}}n_{a}\Pi_{i=1}^{N}\Theta_{i,a_{i}})$
which is a part of GGQD calculation, we can solve the NP-hard, MAX-k-SAT
problem. The NP-hardness of MAX-k-SAT is in the number of the total
variables $N$, which is the number of the measurement angles in the
corresponding GGQD calculation. Thus, the GGQD calculation is NP-hard in the
number of qubits as well.

As for the (non-geometric) global quantum discord calculation (Eq.(\ref%
{eq:global quantum discord})), we did not get the conclusive proof that its
calculation is NP-hard. Nonetheless, we suggest that the injective tensor
norm calculation is also NP-hard as the NP-hardness essentially comes from
the exponentially large dimension of the space of the measurement angles and
the global quantum discord calculation involves $log$ functions which is
generally more difficult to calculate than the polynomials which appear in
the calculation of the GGQD.

\subsection{Two Qubits}

For the case of two qubits, the Bloch vector has two indices and thus can be
viewed as matrix. The GQD can be efficiently and exactly calculated via
singular value decomposition (SVD). 
\begin{eqnarray}
& \rho=\frac{1}{4}(I+\sum_{i\neq0}n_{i0}\sigma_{i}\otimes I+\sum_{j\neq
0}n_{0j}I\otimes\sigma_{j}+\sum_{i,j\neq0}n_{ij}\sigma_{i}\otimes\sigma _{j})
\nonumber \\
& =\frac{1}{4}(I+\sum_{i\neq0}n_{i0}\sigma_{i}\otimes
I+\sum_{j\neq0}n_{0j}I\otimes\sigma_{j}
+\sum_{i,j\neq0,a=1,2,3}\sigma_{i}\otimes\sigma_{j}R_{1,ia}R_{2,ja}d_{a}) 
\nonumber \\
& =\frac{1}{4}(I+\sum_{i\neq0,a}R_{1,ai}n_{i0}\sigma_{a}^{\prime}\otimes
I+\sum_{j\neq0,a}R_{2,aj}n_{0j}I\otimes\sigma_{a}^{\prime\prime}+%
\sum_{a}d_{a}\sigma_{a}^{\prime}\otimes\sigma_{a}^{\prime\prime})
\end{eqnarray}
where $\vec{\sigma}^{\prime}\equiv U_{R_{1}}\vec{\sigma}U_{R_{1}}^{\dagger },%
\vec{\sigma}^{\prime\prime}\equiv U_{R_{2}}\vec{\sigma}U_{R_{2}}^{\dagger}$.
This corresponds to a local change of basis via local unitary transformation
so it does not change the correlation measure. Since we consider the special
case where the marginalized partial states are maximally mixed ($%
n_{i0}=n_{0j}=0,i,j\neq0$), the first two terms are zero, and the state
becomes a Bell diagonal state after SVD is applied. 
\begin{equation}
\rho=\frac{1}{4}(I+\sum_{i,j\neq0}n_{ij}\sigma_{i}\otimes\sigma_{j})
\label{eq:two qubit state with maximally mixed marginal states}
\end{equation}

Now with SVD, $n_{ij}=\delta _{ij}d_{i}$, and the $C$ in Eq.(\ref{eq:global
quantum discord for special case}) is $C=d_{1}\theta _{1,1}\theta
_{2,1}+d_{2}\theta _{1,2}\theta _{2,2}+d_{3}\theta _{1,3}\theta _{2,3},$
ordered so that $d_{1}\geq d_{2}\geq d_{3}$, ($d_{i}\in $ $\mathbb{R}$ and $%
n_{ij}\in \mathbb{R}$). \ The maximum value of $C$ is attained by choosing
the largest $d_{i}$ and setting the corresponding $\theta _{1,i}=1$. This is
much simpler than other methods that have been used \cite{luo_quantum_2008,
ali_quantum_2010}.

With $n_{ij}=\delta _{ij}d_{i}$, the geometric quantum discord as in Eq.(\ref%
{eq:geometric global quantum discord for special case}) now simplifies as 
\begin{eqnarray}  \label{eq:global quantum discord for two qubits with SVD}
D_{GG} & =\frac{1}{4}(d_{1}^{2}+d_{2}^{2}+d_{3}^{2}-\max_{i}d_{i}^{2})=\frac{%
1}{4}(d_{2}^{2}+d_{3}^{2})
\end{eqnarray}
$D_{GG}$ is the sum of the squares of the singular values \textit{excluding}
the largest one \cite{dakic_vedral_brukner_2010,
bellomo_giorgi_galve_franco_compagno_zambrini_2012,
franco_bellomo_maniscalco_compagno_2013}.

The (original) quantum discord is defined as $D=I-J=($total correlation) -
(classical correlation). \ This allows us to interpret this expression for
the geometric quantum discord as follows. The largest singular value of the
Bloch vector quantifies the classical correlation, and the basis vectors
that diagonalize the corresponding density matrix give us the measurement
direction that teases out this correlation. \ If the other directions are
uncorrelated, as they would be in a classical state, then $D=0.$ \ In a
quantum-correlated state, there is residual correlation over and above this
classical correlation in the other directions. \ This additional correlation
is the discord and it is measured by the size of the other singular values.
In other words, the other singular values are how large the off-diagonal
terms of the density matrix would be in the "classical" basis.

The fact that the geometric quantum discord is a function only of the 2nd
and 3rd largest singular values is in sharp contrast to entanglement. For
the Bell diagonal states, the entropy of formation $E$ was calculated can be
computed analytically \cite{lang_quantum_2010}. \ A Bell diagonal state is
parametrized by three parameters $c_{1},c_{2},c_{3}$, whose physically
allowed values (by positivity arguments) form a tetrahedron and the four
corners of the tetrahedron are the maximally entangled Bell states $\psi
^{+},\psi ^{-},\phi ^{+},\phi ^{-}$ given by $%
(c_{1},c_{2},c_{3})_{Bell}=(1,1,1),(1,-1,-1),(-1,1,-1),(-1,-1,1).$ $E$ is a
monotonic function of the distance to the closest corner $%
E(c_{1},c_{2},c_{3})=(c_{1},c_{2},c_{3})\cdot (c_{1},c_{2},c_{3})_{Bell}$.
Unlike geometric quantum discord, entanglement depends mostly on the radius
(distance from the fully mixed state). \ It depends only weakly on the
direction in the vector space of the $c_{i}$. \ This feature of $E$ is
manifests itself in the possibility of entanglement sudden death which is
due to the fact that there exists a finite radius within which all states
are separable and the entanglement is zero \cite%
{schack_explicit_2000,braunstein_separability_1999,gurvits_separable_2003,zhou_disappearance_2011}%
. For Bell diagonal states, an octahedron of separable states resides inside
the tetrahedron, inside which the state is separable\cite{lang_quantum_2010}%
. \ Thus, $E$ is more fragile: more sensitive to external noise.

\subsection{N Qubits}

Turning to the $N$-qubit case, Tucker decomposition can be applied to the
tensor $n.$ For example, for three indices, $n$ may be written as $%
n_{ijk}=\sum_{abc}D_{abc}R_{ai}^{(1)}R_{bj}^{(2)}R_{ck}^{(3)}$, where $%
R^{(\alpha)}\in$ SO(3) but $D_{abc}$ is not in any sense diagonal. No simple
criterion seems to be available to determine whether $D_{abc}$ is of the
form $D_{abc}=D_{a}\delta_{ab}\delta_{bc}$, in which case the decomposition
is called higher order singular value decomposition (HOSVD)\cite%
{bergqvist_higher-order_2010, de_lathauwer_multilinear_2000}. For more than
three qubits, HOSVD would be $n_{i_{1}\cdots
i_{N}}=\sum_{a}D_{a}\Pi_{k=1}^{N}R_{ai_{k}}^{(k)}$. In this case, we have
three principal values and the GGQD can be calculated as in the two qubit
case, as in Eq.(\ref{eq:global quantum discord for two qubits with SVD}).

\section{Sudden transitions of classical and quantum decoherence}

\subsection{Numerical method}

In this section we will be considering low-dimensional symmetric $n$'s. 
\begin{equation}
\rho=\frac{1}{2^N}(I + \sum_{\vec{a},a_i\neq0}n_{\vec{a}}\sigma_{a_1}\otimes%
\cdots\otimes\sigma_{a_N})
\label{eq:N qubit state with maximally mixed marginal states}
\end{equation}
In these cases, it is reasonable to employ a heuristic algorithm used in
other physical contexts to compute $D.$ \ This algorithm demonstrably works
for some simple cases, but we will apply it without proof. \ We regard $%
-n_{ijk...}$as an interaction energy among classical spins $\vec{\Theta}_{i}$
located at sites labeled by $i.$ \ The method starts with a trial set $%
\left\{ \vec{\Theta}_{i}^{\left( 0\right) }\right\} ,$ and uses it to
compute a mean field on each of the sites. \ $\left\{ \vec{\Theta}%
_{i}^{\left( 1\right) }\right\} $ is computed by minimizing the energy at
each site $i$ individually. \ $\left\{ \vec{\Theta}_{i}^{\left( 1\right)
}\right\} $ is used to compute a new mean field that determines $\left\{ 
\vec{\Theta}_{i}^{\left( 2\right) }\right\} $ and the iteration is repeated
to convergence. \ In practice, the convergence may be improved by
introducing a damping factor $\alpha$ with $0<\alpha<1$ and computing $%
\left\{ \vec{\Theta}_{i}^{\left( j\right) }\right\} ^{\prime},$ where 
\begin{eqnarray*}
\left\{ \vec{\Theta}_{i}^{\left( 0\right) }\right\} ^{\prime} & =\left\{ 
\vec{\Theta}_{i}^{\left( 0\right) }\right\} , \\
\left\{ \vec{\Theta}_{i}^{\left( 1\right) }\right\} ^{\prime} & =\left(
1-\alpha\right) \left\{ \vec{\Theta}_{i}^{\left( 0\right) }\right\}
+\alpha\left\{ \vec{\Theta}_{i}^{\left( 1\right) }\right\} \\
\left\{ \vec{\Theta}_{i}^{\left( 2\right) }\right\} ^{\prime} & =\left(
1-\alpha\right) \left\{ \vec{\Theta}_{i}^{\left( 1\right) }\right\}
^{\prime}+\alpha\left\{ \vec{\Theta}_{i}^{\left( 2\right) }\right\} , \\
& ...
\end{eqnarray*}
where $\left\{ \vec{\Theta}_{i}^{\left( 2\right) }\right\} $ is computed by
minimizing the energy in the mean field of $\left\{ \vec{\Theta}_{i}^{\left(
1\right) }\right\} ^{\prime},$ not $\left\{ \vec{\Theta}_{i}^{\left(
1\right) }\right\} .$ \ This method has been used for spin glasses and it is
analogous to calculation methods that have been used for geometric
entanglement \cite{streltsov_simple_2011}.

As a simple example, consider $n=(n_{11},n_{22},...)=(1,\frac{1}{2},0,...)$
with initial guess $\vec{\Theta}_{1}^{(0)}=\vec{\Theta}_{2}^{(0)}=(\frac {1}{%
\sqrt{2}},\frac{1}{\sqrt{2}},0)$, during the first iteration, update $\vec{%
\Theta}_{1}$ to $\vec{\Theta}_{1}^{(1)}=n \cdot \vec{\Theta}%
_{2}^{(0)}=(n_{11}~\vec{\Theta}_{2,1}^{(0)}+n_{12}~\vec{\Theta }%
_{2,2}^{(0)}+\cdots,~n_{21}~\vec{\Theta}_{2,1}^{(0)}+\cdots,~\cdots)$

$=(\frac{1}{2\sqrt{2}},\frac{1}{4\sqrt{2}},0)\rightarrow(\frac{2}{\sqrt{5}},%
\frac{1}{\sqrt{5}},0)$ (normalized), update $\vec{\Theta}_{2}$ to $\vec{%
\Theta}_{2}^{(1)}=n \cdot\vec{\Theta}_{1}^{(1)}=(n_{11}~\vec{\Theta}%
_{1,1}^{(1)}+n_{12}~\vec{\Theta}_{1,2}^{(1)}+\cdots,~n_{21}\vec{\Theta}%
_{1,1}^{(1)}+\cdots,\cdots)=(\frac{1}{\sqrt{5}},\frac{1}{4\sqrt{5}}%
,0)\rightarrow(\frac{4}{\sqrt{17}},\frac{1}{\sqrt{17}},0)$ (normalized). We
can see $\vec{\Theta}_{1}^{(n)},\vec{\Theta}_{2}^{(n)}$ gradually move to
the optimal values of (1,0,0), (1,0,0) as $n$ increases.

This mean-field method is not as accurate or as general as, say, the branch
and bound method\cite{thoft-christensen_branch-and-bound_1986}. \ Indeed,
the general problem is very similar to the calculation of the ground state
of a spin glass, and it is unlikely that mean-field approaches will be
effective. \ For the relatively symmetrical cases we will consider it seems
to be adequate. \ 

\subsection{Dynamics of Discord}

Having achieved some insight into the geometrical structure of quantum
discord, we can make some statements about its dynamics. \ Sudden death of
quantum discord does not occur in a generic system evolution because the
concordant space (where the quantum discord is zero) has measure zero and is
nowhere dense\cite{ferraro_almost_2010, nguyen_topology_2013}. \ Thus a
state picked out at random has positive discord with very high probability
and an arbitrarily small perturbation\ can always be found that will take a
state with zero discord to a state of strictly positive discord. \ By the
same token, the phenomenon of frozen geometric discord (in which the
geometric discord is constant for a finite interval of time) is also not
generic. \ It occurs in highly symmetric situations when the trajectory in
state space parallels the nearest surface of the concordant set.

The stronger anisotropy in the state space (as it depends only on the 2nd
and 3rd singular value as shown in the previous section) of the quantum
discord as compared to the entanglement implies that it can be less
sensitive to decoherence than entanglement, if the decoherence takes the
trajectory in the proper direction. \ We now consider these effects as
manifested in the observation of sudden transitions in the quantum discord.

The type of sudden transition we consider is when the quantum discord $%
D_{GG}\left( t\right) $ decays at first slowly (classical decoherence) until
a certain time $t_{c}$ when it begins to decay more quickly (quantum
decoherence), the transition point being defined by a discontinuous change
of the derivative $dD_{GG}\left( t\right) /dt$ at $t=$ $t_{c}$. The sudden
transition for two qubits was demonstrated for states of the form: $\rho =%
\frac{1}{4}(I+n_{11}\sigma _{11}+n_{22}\sigma _{22}+n_{33}\sigma _{33})$,
and a dynamical model that included only phase flips: $\rho \rightarrow
p\rho +(1-p)\sigma _{z}\rho \sigma _{z},$ where $p\left( t\right) $ is a
monotonically increasing function of time. \ \ Under these circumstances the 
$z$-like components of the generalized Bloch vector do not decay, while the $%
x$- and $y$-like components elements do decay. \ This suggests that the
observation of the transition is connected with the existence of
decoherence-free subspaces.

\begin{figure}[tbp]
\subfigure[Discontinuous derivative] {%
\includegraphics[scale=0.8]{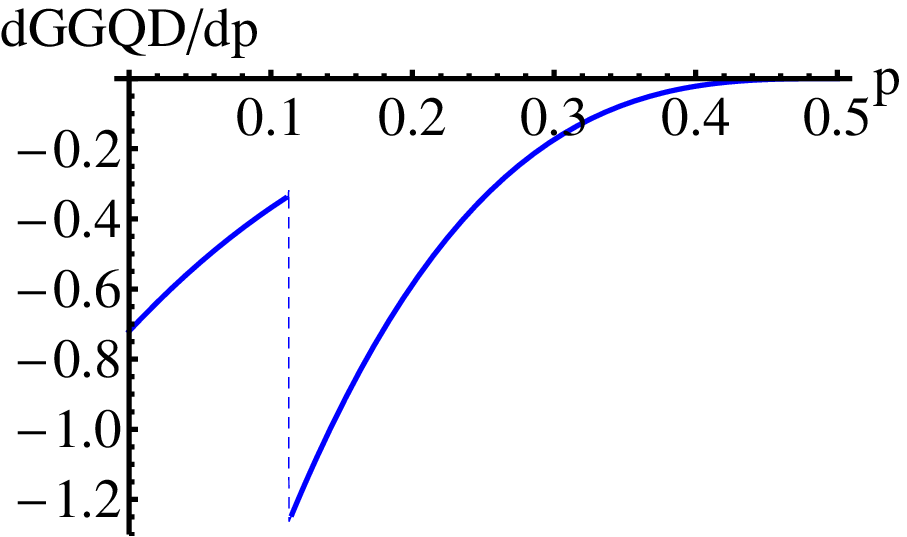}} %
\subfigure[Level crossing] {%
\includegraphics[scale=0.8]{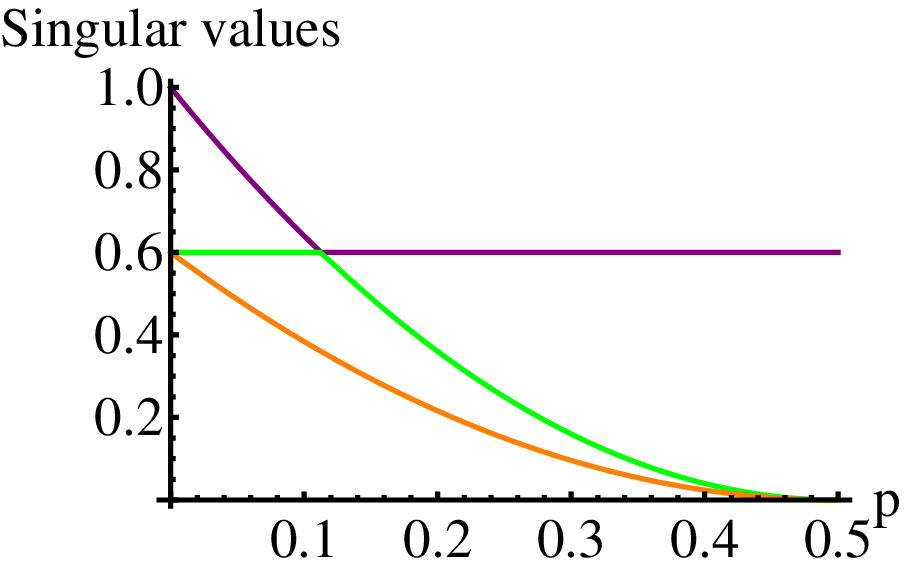}}
\caption{(a) The sudden transition is characterized by a discontinuous
change in the time derivative of the discord. \ Here this phenomenon is
shown for a model in which decoherence comes from phase flips. \ p is the
probability that a flip has taken place, so p is a monotonically increasing
function of time.\ (b) The transition may be traced back to the a level
crossing of singular values f the density matrix. \ Here we show the three
largest singular values as a function of time.}
\label{figure:evolution-sudden}
\end{figure}

It was shown that the condition of the sudden transition with quantum
discord for the state of the form $\rho =\frac{1}{4}(I+n_{11}\sigma
_{11}+n_{22}\sigma _{22}+n_{33}\sigma _{33})$ is either $|n_{11}(p=0)|\geq
|n_{33}(p=0)|$ or $|n_{22}(p=0)|\geq |n_{33}(p=0)|$\cite%
{maziero_classical_2009}. We have the identical condition for geometric
global quantum discord as is clearly seen from Eq.(\ref{eq:global quantum
discord for two qubits with SVD}). The discontinuous change in the slope of
geometric global quantum discord occurs if and only if the first $\left(
d_{1}\right) $ and second largest $\left( d_{2}\right) $ singular values
cross as a function of $p.$ \ In this case $d_{1}$ has a discontinuous
derivative as a function of $p$ (or $t).$ The discontinuity in the derivative of GGQD and its connection with the behavior of the singular values is shown in Figure.\ref{figure:evolution-sudden}. \ The crossing is observed because
the $n_{33}$, which is supposed to be either the second or the third largest
initial singular value, does not change due to the aforementioned symmetry
of the system. \ \ The same behavior occurs for a state of the form: $\rho =%
\frac{1}{4}(I+\sum_{i,j=1,2}n_{ij}\sigma _{ij}+n_{33}\sigma _{33})$ where
the discontinuous change of derivative is also observed.

\begin{figure}[t]
%ptb
\subfigure[Geometric Global QD] {%
\includegraphics[scale=0.8]{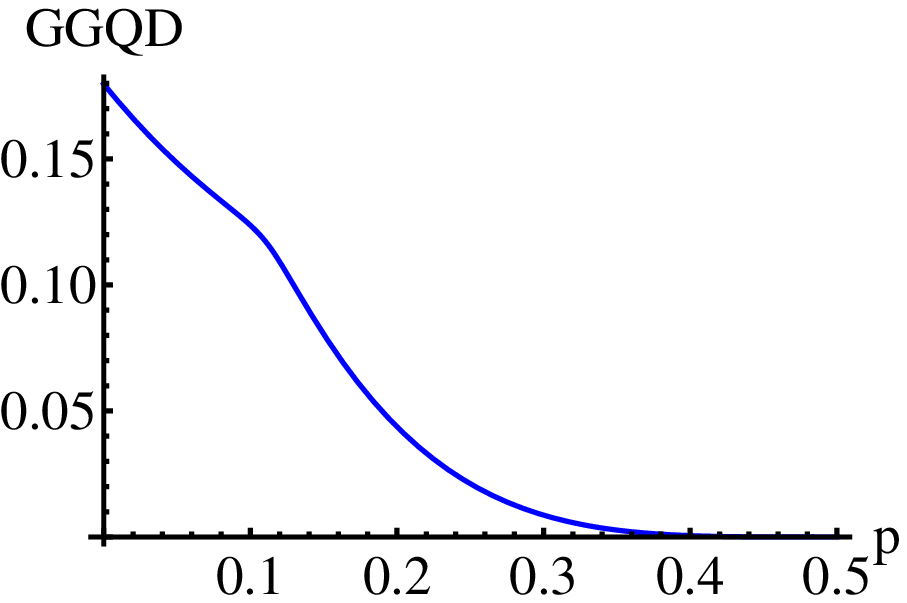}} %
\subfigure[Singular values] {%
\includegraphics[scale=0.8]{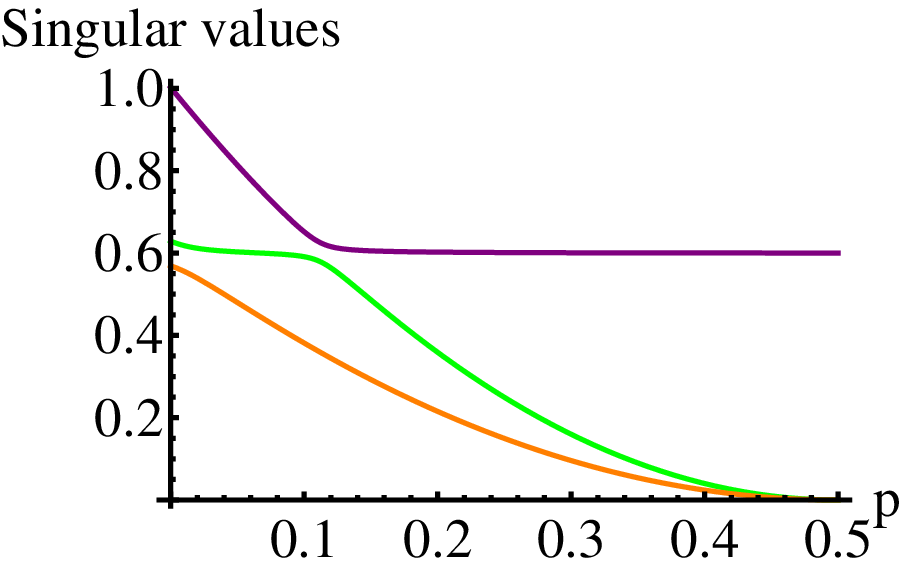}}
\caption{(a) The sudden transition is smoothed when the level crossing is
avoided. \ Here the discord as a function of time shows a smooth crossover
from slow to fast decay. \ \ (b) The three largest singular values of the
density matrix. The crossing of the two largest singular values is now
avoided. \ The symmetry that allowed the level crossing in Figure 1(b) has
been removed. \ }
\label{figure:evolution-Smoothed}
\end{figure}

However, when we have non zero $n_{i3}$, $n_{3i}$ ($i=1,2$) elements, the
symmetry is violated and the level crossing goes away. \ The singular value
evolution is completely smooth and the crossing does not occur. \ The
contrast between the nearly discontinuous and the smooth behavior is shown
in Figure.\ref{figure:evolution-Smoothed}. 
Thus the qualitative conclusion is that the sudden transition is due to a
level crossing, which is essentially always a consequence of symmetry; level
repulsion due to a small symmetry breaking smooths the crossing, giving rise
to a smooth but rapid change, and generic level movement without any
symmetry destroys the transition entirely. \ Similar conclusions have been
reached in \cite{yu_li_fan_2013}. It is important to note that if entropic
definitions of discord are used, then rapid changes may still be observed,
but the behavior of the discord is always continuous, as pointed out in \cite%
{pinto_karpat_fanchini_2013}. This allows us to formulate more
quantitatively the conditions for the observation of the sudden transition,
for which sudden but not discontinuous change in the slope of geometric
global quantum discord is observed, i.e., the level crossing is avoided, but
the gap is small. It is motivated by observing that each singular value has
contributions from various parts of the generalized Bloch vector, of which
only the one from the protected part $n_{33}$ is preserved: $%
d_{a}=\sum_{ij}n_{ij}R_{ia}^{(1)}R_{ja}^{(2)}=n_{\alpha \alpha }R_{\alpha
a}^{(1)}R_{\alpha a}^{(2)}+\sum_{(i,j)\neq (\alpha ,\alpha
)}n_{ij}R_{ia}^{(1)}R_{ja}^{(2)}$, where $d_{a,robust}\equiv n_{\alpha
\alpha }R_{\alpha a}^{(1)}R_{\alpha a}^{(2)}$ is the contribution to $d_{a}$
from the conserved part of Bloch vector. Contributions from other parts
vanish as decoherence proceeds (with $p$ approaching $1/2$). Therefore, we
model the behavior of each singular value $d_{a}$ to monotonically decrease
to $d_{a,robust}$ and the heuristic argument for the sudden transition is
that the largest singular value crosses a smaller one as $p$ increases. In
this picture, the criteria for the crossing of the largest singular value
and the one of the smaller ones is either $%
|d_{2,robust}(p=0)|>|d_{1,robust}(p=0)|$ or $%
|d_{3,robust}(p=0)|>|d_{1,robust}(p=0)|$. But this model does not accurately
describe the actual behavior, as it does not capture that the $R^{(i)}$s
change over time as well and the only one singular value has nonzero value
of $n_{33}$ at full decoherence ($p=1/2$). However, if the size of the
off-diagonal terms are small enough compared to the other parts, above
argument still gives reasonable condition for sudden transition and it
converges to the accurate condition as $\epsilon \rightarrow 0$. The
previously shown condition of \cite{maziero_classical_2009} is a special
case for $\epsilon \rightarrow 0$ case where $n_{ij}$ is diagonal. 
%For state of the form of eq(\ref{eq:two qubit state with maximally mixed marginal states}), there are essentially three quantities (three singular values) to store correlation. One strategy to conserve the quantum correlation in terms of global geometrc quantum discord is to avoid preserving the largest value, as the largest value is irrelavant to geometric global quantum discord.

In order to estimate the chance of observing the sudden transition from
arbitrary states of the form of Eq.(\ref{eq:two qubit state with maximally
mixed marginal states}), we assume the axis of each rotation $R^{(i)}$ is
uniformly distributed over the unit spheres: Pr$(\theta =\theta _{0},\phi
=\phi _{0})~d\theta ~d\phi =\frac{1}{4\pi }\sin (\theta _{0})~d\theta ~d\phi 
$. It corresponds to the random choice of SU(2) unitary matrix according to
Haar measure, and we choose the state at random given the three singular
values $d_{a},a=1,2,3$. The rotation matrix without nonzero off-diagonal
entries corresponds to a rotation in two dimensional space, and the volume
of the sets of 2-dimensional rotation matrices in the space of 3-dimentional
rotation matrices of course has zero measure. Even if we loosen the
condition by allowing a small deviation from the 2-dimensional rotation
which results in small probability $\epsilon \ll 1$ for the rotation matrix
to have the desired character, the probability of the sudden transition for
the 2-qubit system is proportional to $\epsilon ^{2}$. For the three or more
qubits case where subsystems are maximally mixed ($\rho =\frac{1}{2^{N}}%
(I+\sum_{a_{1}=1}^{3}\cdots \sum_{a_{N}=1}^{3}n_{\vec{a}}\sigma _{\vec{a}})$%
) and HOSVD can be applied to the generalized Bloch vector as mentioned
earlier ($n_{i_{1}\cdots i_{N}}=\sum_{a}d_{a}\Pi _{k=1}^{N}R_{ai_{k}}^{(k)}$%
), each rotation matrix $R^{(i)}$ independently gives a additional factor of 
$\epsilon $ to the chance of the sudden transition so the chance of the
transition decays exponentially as $\epsilon ^{N}$, as $N,$ the number of
qubits, increases.

\section{Conclusion}

We proposed the use of the generalized Bloch vector for calculation of
quantum discord. It makes calculation easier for previously known cases and
provides some useful insights on quantum correlation. We showed (under
certain weak assumptions) that the calculation of the geometric global
quantum discord is an NP-hard problem by considering a certain interesting
class of multi-qubit states. It appears to be significantly more difficult
to prove corresponding statements for other, non-geometric measures of
discord, since they involve more complicated functions.

For states of the form of Eq.(\ref{eq:N qubit state with maximally mixed
marginal states}), we suggested a numerical method to calculate geometric
global quantum discord, which appears to be give good results for many
interesting models. \ When higher-order singular value decomposition is
applicable, we proposed a condition to observe the sudden transitions in the
geometric global quantum discord, assuming the part preserved by the
symmetry of the system and the other parts do not mix significantly. For
randomly chosen states, the sharp sudden transition has only a small chance
of being observed in the 2-qubit case and it becomes exponentially rarer as
the number of qubits increases, because the number of restrictive symmetry
conditions needed for this phenomenon to occur increases rapidly with system
size.

\ack
We thank Nga Nguyen for informative discussions. 
This project was supported by DARPA-QuEst (MSN118850) and the U.S. Army Research Office
(W911NF-08-1-0482, W911NF-12-1-0607). The views and conclusions contained in
this document are those of the authors and should not be interpreted as
representing the official policies, either expressly or implied, of the US
Government.

\section*{References}

%\bibliography{discordWriting_revised}
\providecommand{\newblock}{}

\end{document}